\documentclass{svjour2}                    
\smartqed  
\usepackage{graphicx}
\begin{document}

\title{Influence of Ortho-H$_2$ Clusters on the
Mechanical Properties of Solid Para-H$_2$\thanks{Support provided by
NSF Grants DMR 0207071 and 0706339.} }


\author{A. C. Clark \and Z. G. Cheng \and M. Bowne \and X. Lin  \and M. H. W. Chan
}


\institute{A. C. Clark \and Z. G. Cheng \and M. Bowne \and X. Lin
            \and M. H. W. Chan \at
              Department of Physics, The Pennsylvania State University,
              University Park, Pennsylvania 16802, USA  \\
            \and
            A. C. Clark \at
              \email{cctony1@gmail.com}             \\
              \emph{Present address:} Department of Physics, University of Basel, CH-4056 Basel, Switzerland \\
            \and
            M. Bowne \at
              \emph{Present address:} FRABA Inc., Hamilton, New Jersey 08609, USA  \\
            \and
            X. Lin \at
              \emph{Present address:} Department of Physics, Massachusetts Institute of Technology, Cambridge, Massachusetts 02139, USA \\
}

\date{Received: date / Accepted: date}

\maketitle

\begin{abstract}
The quantum diffusion of ortho-H$_2$ impurities in solid para-H$_2$
has been investigated with a torsional oscillator technique. The
unclustering dynamics of agglomerated impurities with increasing
temperature are found to greatly affect the resonant period of
oscillation. Based on the observed trends in both experimental data
and complementary finite element calculations, we find that the
oscillation period predominantly reflects changes in the solid's
elastic response to the minute inertial stresses imposed on it. It
is likely that the small but abrupt increase in the resonant period
observed above 100 mK in high purity H$_2$ samples is due to the
interplay between remnant ortho-H$_2$ impurities and dislocations,
and not a signature of a supersolid to normal solid transition.

\end{abstract}

\section{Introduction}
\label{intro}

The condensed hydrogen isotopes belong to a unique class of solids,
quantum crystals, having constituent particles that exhibit
extremely large zero point fluctuations about their equilibrium
lattice sites. For the lightest isotope, H$_2$, this motion can be
as large as 18\% the nearest neighbor distance \cite{nielsen}, well
above the value considered as a melting criterion for most
materials. Consequently, even at zero temperature there is a finite
overlap of particle wavefunctions between lattice sites. The concept
that zero point motion allows crystalline defects to become
delocalized at low temperature, and the conjecture that this could
result in the coexistence of elasticity and superfluidity
\cite{andreev}, launched a series of experiments on both the solid
hydrogens and (the even more quantum mechanical) helium
\cite{meisel}.

At present there is still no indication of such a ``supersolid''
phase existing in H$_2$, and only recently has there been any
evidence for such a phase in $^4$He \cite{nature}. The majority of
experimental results on $^4$He suggesting the existence of
supersolidity have been carried out using a torsional oscillator
(TO) technique \cite{TOcornell}. The underlying idea is that of an
Andronikashvili experiment, i.e., the response of fluid to an
oscillatory driving force when it is confined to dimensions much
less than the viscous penetration depth. Information about the
superfluid fraction of the liquid can be readily acquired from the
resonant period $\tau$ of the TO based on a simple two fluid model.
The normal fluid is viscously clamped to the cell, while the
inherent zero viscosity of the superfluid component allows it to
decouple from the system, thereby decreasing $\tau$ by an amount
proportional to the superfluid mass. The story is more complicated
for solids since the measured response also depends on the
mechanical properties of the sample \cite{balatsky}. The most severe
implication in regard to $^4$He experiments is the possible lack of
supersolidity altogether. However, a number of experiments and
analysis tend to suggest otherwise
\cite{sci,reppy,kojima,dorsey,prb,kojima2,seamus,joshy}. In one of
our own works \cite{prb}, we used a finite element method (FEM) to
calculate the effect that changes in the solid $^4$He elastic moduli
would have on the TO resonant period. Combining these calculations
with experiment we could extract the increase in the shear modulus
{\it G} necessary to explain the phenomenon. In almost all cases the
enhancements were completely nonphysical, and always found to be
greater than that which has been measured \cite{joshy,day}.
Analogously, it was found that viscoelastic \cite{dorsey} and/or
glassy \cite{seamus} behavior can most probably only account for a
small fraction of the period shifts observed.

In this paper we carry out a similar FEM analysis of data from
earlier TO measurements on solid H$_2$, some of the latter of which
was presented in \cite{h2prl}. Our previous study was part of a
search for supersolidity in H$_2$. Interestingly, we observed an
anomalous resonant period shift below $\sim$100 mK that depended on
the concentration {\it x} of remnant ortho-H$_2$ (o-H$_2$)
impurities. However, this apparent mass flow was found to be
inconsistent with the notion of supersolidity during a blocked
annulus the control experiment (see Sects.~\ref{sec:3} and
\ref{sec:4}) \cite{sci,reppy}.

A plethora of studies exist on defect (impurity) diffusion in
hydrogen crystals in the temperature range 25 mK $<$ {\it T} $<$ 4 K
\cite{silvera,meyer1,meyer2}. The variety of experimental techniques
vary from ``direct'' detection methods such as nuclear magnetic
resonance (NMR) of the different molecular, atomic and ionic spins
in samples, to ``global'' detection of thermodynamic properties that
are not only affected by diffusion itself but also by the final
clustered state of o-H$_2$ that is favorable below 1 K. These latter
studies typically involved monitoring the evolution of pressure,
specific heat, thermal conductivity, etc. following abrupt
temperature changes. Other experiments have concentrated on the
mechanical properties of hydrogen crystals. Anomalous plasticity and
creep of solid hydrogen samples have been observed below $\sim$5 K,
and found to be quite sensitive to crystal quality and purity
\cite{one,two,three,four}. However, these very interesting
investigations have been limited to stresses amplitudes ($\sigma$
$>$ 500 Pa) much higher than in TO measurements (between 20 and 400
mPa in this work) and {\it T} $>$ 1 K, where quantum diffusion of
o-H$_2$ impurities does not lead to clustering. Here we explore
effects of impurities on the oscillatory response of solid H$_2$
under very low stress and in the temperature regime where impurity
clustering enabled by quantum diffusion of spins is significant.

\begin{figure*}
  \includegraphics[width=0.75\textwidth]{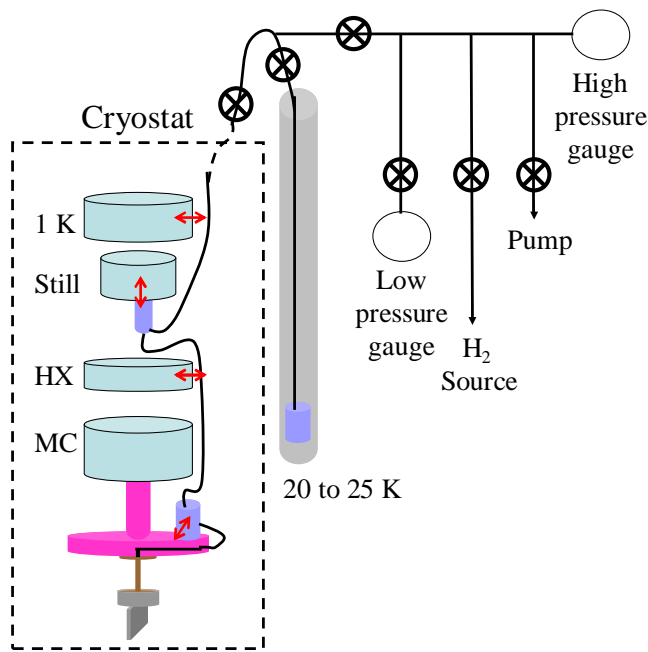}
\caption{Gas handling system that enabled the growth of H$_2$ with
{\it x} $<$ 0.005. The first stage of OP conversion is carried out
at 20 K in a column separate from the dilution refrigerator. H$_2$
is condensed into the cell from the column. The lower conversion
chambers in the capillary system are typically held around 15 K
during sample growth. A final conversion ``filter'' just at the
entrance to the torsion rod was made by filling the capillary within
the bottom stage with FeO(OH).}
\label{fig:1}       
\end{figure*}

\section{Samples}
\label{sec:2}

\begin{figure*}
  \includegraphics[width=0.75\textwidth]{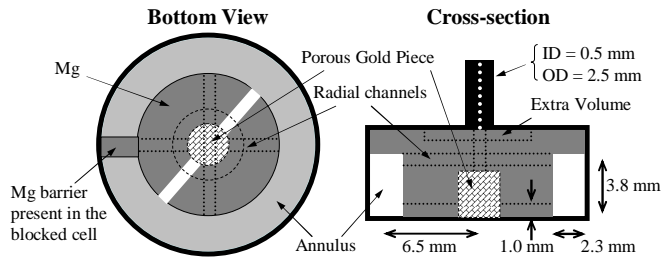}
\caption{Internal geometry of the open and blocked annulus TOs. The
porous gold layer constitutes 2\% of the total cell volume, and
negligible rotational inertia. It was made by leaching Ag out from a
machined AgAu cylinder, which is 2.5 mm in diameter and 3.3 mm tall.
The resonant periods of the open and blocked cell TOs are 709,693 ns
and 703,943 ns, respectively.}
\label{fig:2}       
\end{figure*}

Three different varieties of hydrogen were used in these
experiments: HD-depleted H$_2$ gas, ultra high purity (UHP) H$_2$
gas, and UHP HD gas. The isotopic compositions were {\it n} $<$
10ppm HD and/or D$_2$, 205$\pm$5 ppm HD \cite{mspec} and $<$ 2\%
H$_2$ and/or D$_2$, respectively. The bulk of the measurements were
carried out on HD-depleted samples. A schematic of the capillary
system used to grow samples is given in Fig.~\ref{fig:1}. Drawings
depicting the inner cell of the TOs used in these experiments are
shown in Fig.~\ref{fig:2}.

Apart from isotopic impurities, one can also speak of spin
impurities. The two lowest rotational energy levels of the H$_2$
molecule are the para- and ortho- states, having {\it J}  = 0 and 1,
respectively. The asymmetry of the spin wave function for {\it J} =
1 requires that the spatial wave function also be asymmetric. Thus,
o-H$_2$ impurities distort the surrounding lattice, which is made up
of spherically symmetric p-H$_2$ molecules. Below 4 K the
equilibrium concentration {\it x$_{eq}$} of o-H$_2$ is essentially
zero. However, since the ortho-to-para (OP) conversion rate in the
solid is very slow the most efficient way to produce samples with a
small {\it x} is by exposing the liquid to a high surface area,
magnetic material prior to crystallization. Thus, H$_2$ was first
condensed into a column that housed a 40 cm$^3$ chamber partially
filled with FeO(OH). The entire column was immersed in a storage
dewar of liquid $^4$He. The temperature of the inner chamber was
stable at any fixed height above the surface of liquid $^4$He. Prior
to making solid samples the H$_2$ liquid was kept at 15 K for
several hours or even overnight, at which temperature {\it x$_{eq}$}
$\sim$ 10$^{-4}$. To condense H$_2$ into the cell the temperature of
the column was raised to between 20 and 25 K, where the vapor
pressure is $>$ 1 bar and 10$^{-4}$ $<$ {\it x$_{eq}$} $<$
10$^{-2}$. For HD samples the gas was directly condensed into the
cell.

Since OP conversion is an exothermic process, the average {\it x}
for each sample in the torsion bob was determined by keeping the
mixing chamber at 20 mK and measuring the temperature difference
across the torsion rod. Based on this measurement and the thermal
conductance of the Be-Cu torsion rod, we could infer the power being
emitted during OP conversion, and hence the initial concentration
{\it x}$_0$ (and all subsequent {\it x}). Simply, we used the
relation
\begin{equation}
\frac{\kappa_{BeCu}A\Delta T}{l}=\frac{Ukx_0^2}{(1+kx_0t)^2}
\end{equation}
where the heat of conversion {\it U} = 1.06 kJ mol$^{-1}$, the
reaction rate k = 1.9\% h$^{-1}$, and the thermal conductivity of
BeCu {\it K$_{BeCu}$} was measured in the empty cell and found to
agree extremely well with literature \cite{groger}. The smallest
measurable temperature gradient of $\sim$0.1 mK corresponds to {\it
x} = 0.005.

\section{Experimental Results}
\label{sec:3}

Upon cooling H$_2$ samples to 20 mK, one to two weeks were allowed
for equilibration, during which time $\tau$ dropped smoothly until
finally stabilizing to within a drift of $<$ 0.05 ns per day. The
drift was completely halted by raising the temperature to 40 mK.
After what was considered complete ``equilibration'' at 40 mK, {\it
T} was raised in successive steps, for each of which $\tau$ was
measured as a function of time. When the temperature sweep was
complete the system was returned to 20 mK and allowed to
re-equilibrate prior to a new scan. Since no long relaxation times
were observed for HD samples, the rate of data acquisition was only
limited by thermalization.

\begin{figure*}
  \includegraphics[width=0.75\textwidth]{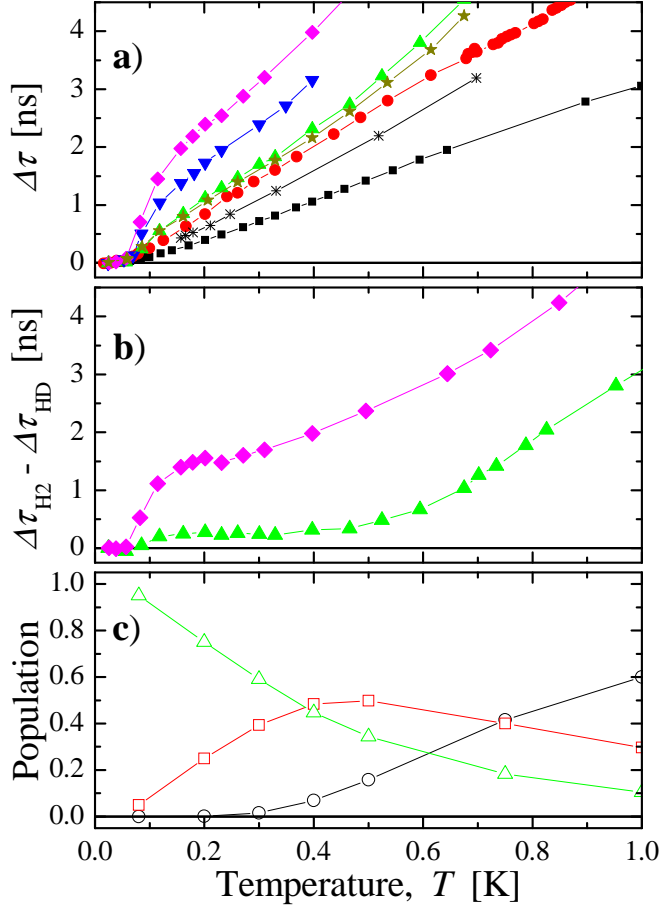}
\caption{a) {\it T}-dependence of $\Delta\tau$ for (black squares)
the empty TO and the cell with (red circles) 91\% filling of HD,
(black asterisks) 88\% filling of H$_2$ with {\it x} = 0.15, (brown
stars) 83\% filling of H$_2$ with {\it x} = 0.025, (green triangles)
83\% filling of H$_2$ with {\it x} = 0.02, (blue inverted triangles)
68\% filling of H$_2$ with {\it x} $<$ 0.005, and (magenta diamonds)
88\% filling of H$_2$ with {\it x} $<$ 0.005. All datasets have been
shifted vertically to match at {\it T} = 0 K for easy comparison. b)
$\Delta\tau$ for (green triangles) 83\% filling of H$_2$ with {\it
x} = 0.02 and (magenta diamonds) 88\% filling of H$_2$ with {\it x}
$<$ 0.005, after subtracting the {\it T}-dependence of $\Delta\tau$
for the HD sample. Although the low temperature feature appears to
be of similar nature to that in solid $^4$He experiments \cite{sci},
it is not due to the onset of superfluidity \cite{h2prl}. c)
Relative population of (black open circles) singles, (red open
squares) pairs and (green open triangles) triangles is shown for
{\it x} = 0.01. The curves are taken from \cite{spt1} and are based
on a simple statistical model.}
\label{fig:3}       
\end{figure*}

Several datasets corresponding to different hydrogen samples of
annular geometry are shown in Fig.~\ref{fig:3}a. All samples were
grown and measured in the TO from \cite{h2prl}. Since we are
interested in changes to $\tau$ following an increase in
temperature, constant offsets have been subtracted from each trace
such that they all coincide at 0 K. The temperature dependence of
the relative period shift $\Delta\tau$ between 150 mK and 1 K is
approximately linear for the empty cell, HD sample, and {\it x} =
0.15 H$_2$ sample. A small deviation from linearity at the lowest
temperatures is typical of BeCu TOs, as is the enhanced {\it
T}-dependence (steeper slope) for TOs containing some amount of
condensed gas. The second effect is not entirely understood, but has
been observed in both solid $^4$He experiments and studies of thin
absorbed $^4$He films \cite{agnolet}. The lowest temperature of 150
mK reached for the {\it x} = 0.15 sample is due to the large OP heat
of conversion continually generated.

There are qualitatively different features found in data from H$_2$
samples with lower ortho-concentration. First, there is a period
shift centered around 100 mK in each sample, which increases in size
with decreasing {\it x}. Second, there is an additional deviation
above 500 mK from the expected linear temperature dependence of
$\Delta\tau$. The third and most striking difference, which cannot
be extracted from Fig.~\ref{fig:3}, is the existence of long
equilibration times at certain temperatures.

\begin{figure*}
  \includegraphics[width=0.75\textwidth]{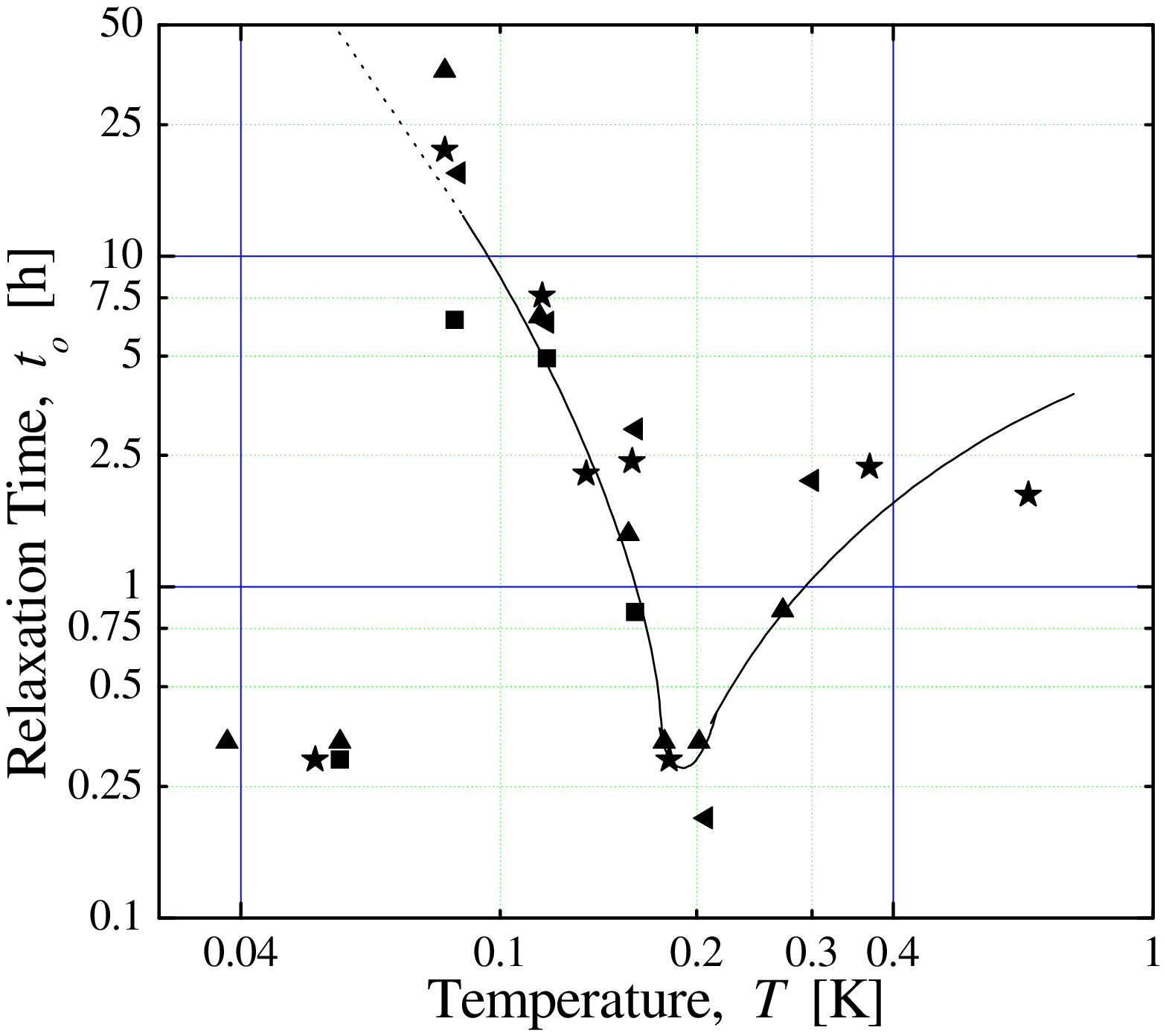}
\caption{{\it T}-dependence of {\it t$_O$} extracted from the time
evolution of $\Delta\tau$. Data are taken from a number of different
samples ({\it x} $<$ 0.025) and stress amplitudes (10 $<$ $\sigma$
$<$ 400 mPa). Below 60 mK the short {\it t$_O$} values measured may
reflect the thermal equilibration time, whereas the true
configurational relaxation times of o-H$_2$ are too long to be
measured. As {\it T} increases from 80 and 180 mK the breakup of
large o-H$_2$ clusters into pairs will occur at faster rates,
shortening the time necessary for equilibration. For {\it T} $>$ 200
mK the picture is further complicated as pairs begin to dissociate.
At high enough temperature {\it t$_O$} will only depend on the
motion of single o-H$_2$ molecules.}
\label{fig:4}       
\end{figure*}

We investigated the temperature dependence of these relaxation times
{\it t$_O$}, which are displayed for a number of samples in
Fig.~\ref{fig:4}. The time dependence of the period was not
exponential in many instances (e.g., see the ``82 mK'' trace in
Fig.~\ref{fig:3} inset of \cite{h2prl}). Since complete
equilibration was obtained for {\it T} $>$ 60 mK we could manually
extract the time necessary to reach the e$^{-1}$ value of each
individual period shift following every temperature step. The
overall behavior was the same for all samples in three different TOs
and for all concentrations in the range studied,
$\sim$5$\times$10$^{-3}$ $<$ {\it x} $<$ 0.025. There was also no
definitive dependence found on oscillation amplitude (i.e., rim
velocity or inertial stress \cite{prb}), which was varied by a
factor of ten. For all samples $\tau$ appeared to equilibrate
quickly for 40 mK $\leq$ {\it T} $\leq$ 60 mK. A more likely,
alternative interpretation is that {\it t$_O$} was so long that the
drift in the period measurement was hidden in the noise, and that
true thermodynamic equilibrium was not achieved for {\it T} $<$ 60
mK. Increasing the temperature further (above 60 mK) revealed
extremely slow relaxation of $\tau$. Near and above 180 mK it was
found that the {\it t$_O$} shortened to less than the thermal
equilibration time. Slow relaxation resumed at {\it T} $>$ 500 mK,
but data was not obtained close enough to equilibrium to determine
{\it t$_O$}. An approximate lower limit is $\sim$5 h.

\begin{figure*}
  \includegraphics[width=0.75\textwidth]{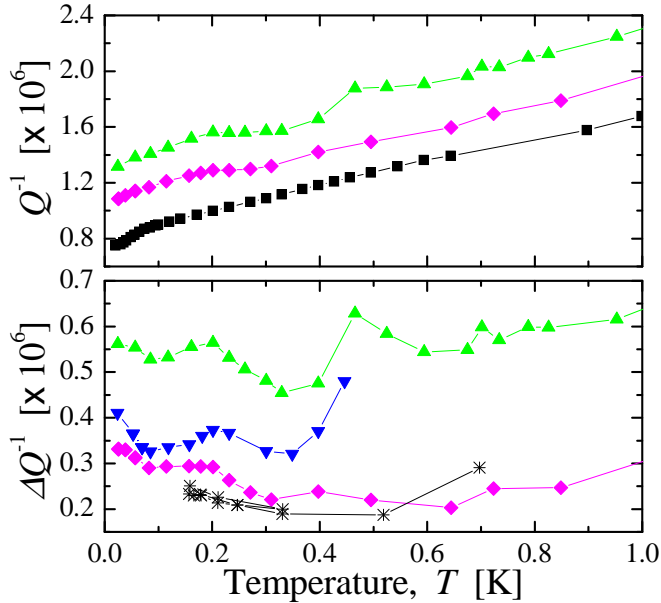}
\caption{Dissipation during oscillation. The legend is identical to
that in Fig.~\ref{fig:3}. In the bottom panel the empty cell
background has been subtracted. The excess dissipation due to H$_2$
depends on {\it T}, increasing over temperature intervals during
which a large degree of cluster dissociation is believed to occur.}
\label{fig:5}       
\end{figure*}

The internal dissipation {\it Q}$^{-1}$ of the TO was also
monitored, simultaneously with the period. Several traces are shown
in the Fig.~\ref{fig:5}. The empty cell background is roughly linear
over the entire temperature range, but exhibits a kink in the slope
near 75 mK, above which the dependence is weaker. The presence of a
H$_2$ sample increases the dissipation in the system. This excess
dissipation $\Delta${\it Q}$^{-1}$ is displayed in the bottom panel
of the figure. The deviations from the empty cell background, which
are more pronounced in samples with {\it x} $<$ 0.15, will be
discussed in Sect.~\ref{sec:5}.

Finally, in our earlier work \cite{h2prl} a blocked annulus control
experiment was carried out \cite{sci,reppy} in order to ascertain
whether the observed effect in solid H$_2$ is a signature of
supersolidity. The purpose of the barrier in the annulus is to make
the path of irrotational superflow much more tortuous, thus leading
to a predictable decrement in the measurable supersolid fraction
\cite{fetter}. We observed very similar $\Delta\tau$ and {\it t$_O$}
in the blocked annulus, clearly indicated that the period shift in
H$_2$ is not related to supersolidity.

\section{FEM Calculations}
\label{sec:4}

Since the internal geometries of the two TOs (see Fig.~\ref{fig:2})
from \cite{h2prl} are more complicated than in $^4$He experiments
\cite{prb}, using precise cell dimensions resulted in very large
numbers of elements and made simulation times unfeasible. Due to
simplifications that had to be made, most of the analysis should be
considered qualitative.

\begin{figure*}
  \includegraphics[width=0.75\textwidth]{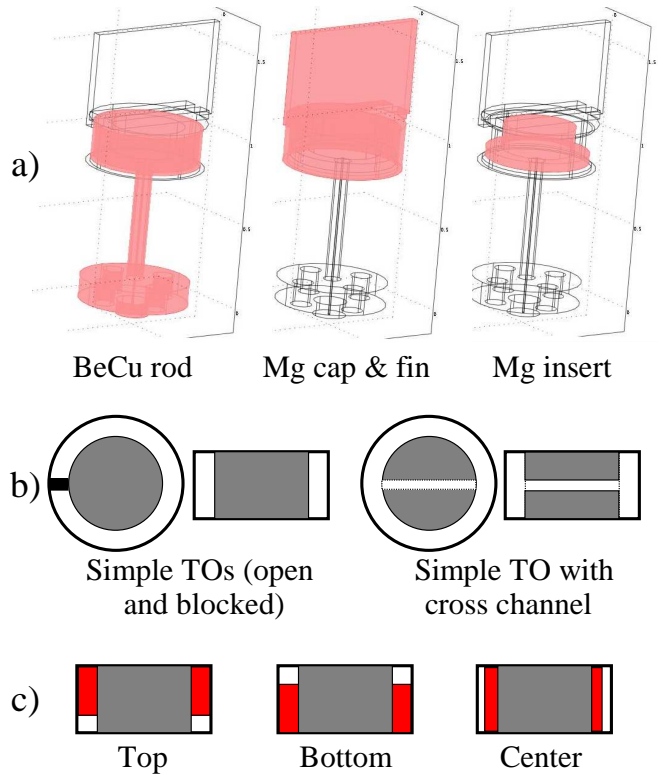}
\caption{a) The three main components making up the open annulus
cell used in the first set of FEM calculations. Several of the
internal features, such as the cross channels and porous gold piece
(see Fig.~\ref{fig:1}), had to be removed. b) Top and side views of
the hydrogen space within the abridged TOs used for most
calculations. The open and blocked annulus TOs were simplified even
further by replacing the Mg insert from (a) with a Mg cylinder. The
simplified open annulus was made slightly more complex in one series
of calculations by adding a cross channel. To keep $\tau$ fixed
around 700,000 ns for all TOs (and matched to that in (a)) the outer
Mg cap and fin were adjusted in size. c) Schematic of the three
different distributions of H$_2$ within the cell that were modeled.
This effect was investigated using the annulus cell from (a).}
\label{fig:6}       
\end{figure*}

In Fig.~\ref{fig:6}a we highlight the three key components of the
open annulus TO for the most geometrically accurate FEM model. It
was necessary to neglect the filling channels and porous gold core
(the latter of which was included in order to lower the
hydrogen-metal thermal boundary resistance). Since the insertion of
a block in the annulus also increased the element number above our
simulation ability, FEM calculations using even simpler internal
geometries were necessary (see Fig.~\ref{fig:6}b). The geometry of
the TOs used in these calculations were designed to have the same
annular dimensions and similar resonant periods to the experimental
values. One cross channel (see Fig.~\ref{fig:6}b) was also included
in some calculations. In addition to altering the dimensions for the
metal components of the TOs, we also investigated the effects from
three different H$_2$ filling geometries: top, bottom and center
(see Fig.~\ref{fig:6}c). These configurations were used for three
different filling factors (i.e., period shifts due to different
samples normalized by the maximum possible H$_2$ mass loading):
50\%, 80\% and 90\%.

The FEM results are listed in Tables~\ref{tab:1} and \ref{tab:2}.
The primary observations are the following: 1) calculated period
shifts $\Delta\tau_{FEM}$ caused by decreases in the shear modulus
of H$_2$ do not depend strongly on the distribution of H$_2$ within
the cell, but rather on the sample filling factor (this has also
been observed experimentally); 2) these same shifts are almost
completely unaffected by the presence of a cross channel; 3) the
shift is essentially the same for open and blocked annulus
geometries; 4) changes to the bulk modulus {\it K} have little or no
effect on both the open and blocked annulus TOs.

\section{Discussion}
\label{sec:5}

The long relaxation times displayed in Fig.~\ref{fig:4} exist at
similar temperatures and are of similar magnitude to those from NMR
studies of dilute ortho-para mixtures, which have been identified
with the clustering of o-H$_2$ molecules (mostly into pairs)
following a sharp drop in {\it T} \cite{meyer1,meyer2}. The
impurities effectively travel throughout the lattice via the
transfer of {\t J} from molecule to molecule rather than by particle
exchange, thus it is spin (rather than mass) diffusion. This process
is called resonant OP conversion \cite{oyarzun}. Similar timescales
are also observed for the restoration of isolated singles when pairs
breakup during subsequent warming of the sample. On the contrary,
Schweizer {\it et al.} found growth and decay times of o-H$_2$ pairs
to exhibit considerable hysteresis when crystals were held below
$\sim$50 mK for 10 h or more \cite{II}. Since diffusion rates for
both singles and pairs are enhanced below 100 mK formation of larger
clusters will occur. These triangles, etc. are essentially immobile
since they cannot propagate by the same resonant OP mechanism
responsible for single and pair diffusion \cite{kran}. Once such a
configuration has formed it appears, based on the lack of any
observable pairs or singles, to remain frozen in for days even upon
warming to 300 mK \cite{II}. The rate of pair formation increases at
higher {\it T}. For example, the presence of pairs was detected by
Schweizer {\it et al.} following a 40 h ``anneal'' at 600 mK.

It is therefore important to point out the difference of the present
investigation from the majority of earlier works. In previous
studies the initial state was prepared at high temperature where the
number of singles dominates over pairs and larger clusters
\cite{spt1,spt2}. Here, clustering is allowed to ensue for up to two
weeks at 20 mK so that, according to decay times in the literature,
no isolated o-H$_2$ singles \cite{li} or pairs \cite{III} remain in
our samples at the beginning of each {\it T} sweep. After some weeks
the overall configuration of o-H$_2$ will not change significantly
since triangles and larger clusters cannot propagate. As the
temperature of the TO is increased it is then possible to monitor
the unclustering of these large agglomerations.

Before proceeding, it should be stressed that the short
equilibration times in HD samples and the {\it x} = 0.15 H$_2$
sample \cite{noteJarvis}, as well as the lack of dependence of {\it
t$_O$} on {\it x}, are all consistent with our expectations if the
phenomenon observed in dilute ortho-para mixtures is related to the
unclustering and diffusion of spin impurities. Since these processes
clearly affect the resonant period of the TO, an obvious question
must be addressed: what physical property of hydrogen does the TO
measure? In \cite{h2prl} we briefly discussed a few possibilities.
For example, the wetting properties of H$_2$ \cite{wetting} could be
influenced by the local OP composition. It is difficult to predict
what consequences this would have on the resonant period
measurement. However, in the case of slippage occuring between the
sample and cell walls one would expect corresponding period shifts
to increase dramatically with oscillation amplitude and eventually
become entirely irreproducible upon cycling, contrary to our
observations.

We also pointed out that the moment of inertia is sensitive to the
radial density profile of a H$_2$ sample, and thus rearrangement of
impurities within the cell could reduce or enhance the resonant
period. Since the density of o-H$_2$ is only 1.7\% higher than that
of p-H$_2$ \cite{silvera}, ortho-concentrations less than a few
percent cannot influence the moment of inertia of H$_2$ (by
clustering at small or large radii within the cell) by more than a
few parts in 10$^4$ (and thus the period by a few parts in 10$^9$).
Although this is similar to the observed $\Delta\tau$ it is an
extreme case that is clearly nonphysical in view of tunneling
probabilities of triangles and larger clusters \cite{kran}.

\begin{table}
\caption{FEM results for different TOs. To investigate the possible
correlation between o-H$_2$ dynamics and the mechanical properties
of the solid, $\tau$ was calculated for different values of the
shear and bulk moduli. The period shifts, listed below (in
nanoseconds) for different TOs, correspond to decreases in {\it G}
or {\it K} by 10\%.}
\label{tab:1}       
\begin{tabular}{ccccc}
\hline\noalign{\smallskip}
Modulus & Annulus & Simple Annulus & Simple Block & Simple Channel \\
\noalign{\smallskip}\hline\noalign{\smallskip}
{\it G} & 0.164 & 0.716 & 0.697 & 0.733 \\
{\it K} & -- & 0.028 & 0.028 & -- \\
\noalign{\smallskip}\hline
\end{tabular}
\end{table}

\begin{table}
\caption{The effect of hydrogen distribution within the cell.
$\Delta\tau_{FEM}$ were calculated for different sample filling
factors and configurations (see Fig.~\ref{fig:6}c). The period
shifts, in nanoseconds, correspond to a 10\% decrease in {\it G}.}
\label{tab:2}       
\begin{tabular}{ccccc}
\hline\noalign{\smallskip}
Filling Factor & Top & Bottom & Center & $\Delta\tau_{FEM}$[Bottom]/$\Delta\tau_{FEM}$[Full]\\
\noalign{\smallskip}\hline\noalign{\smallskip}
50\% & 0.072 & 0.086 & 0.069 & 52.4\% \\
80\% & 0.129 & 0.128 & 0.109 & 78.0\% \\
90\% & -- & 0.133 & -- & 81.1\% \\
\noalign{\smallskip}\hline
\end{tabular}
\end{table}

Since most of the samples did not entirely fill the cell, one can
also consider the thermal contraction of H$_2$. The molar volume
change from the freezing point {\it T$_m$} down to zero temperature,
1.1\% \cite{krupskii}, can decrease the outer radius of a confined
ring of H$_2$ by no more than $\sim$0.37\%, and thus the moment of
inertia by $\sim$2\% for the given cell dimensions. This percentage
of the mass loading sets an upper limit of $\sim$60 ns. For all
hydrogen samples $\tau$ did drop by several tens of nanoseconds
between the freezing point and low {\it T}. However, contraction
should only affect the resonant period between 0.5{\it T$_m$} and
{\it T$_m$}, whereas we observed large changes even when cooling
below 4 K.

Physical quantities that also change significantly below the
freezing point are the bulk and shear moduli, increasing by at least
10\% or 20\% \cite{krupskii,manz,swenson}. Although the majority of
stiffening is expected to occur just below {\it T$_m$}, the pinning
of dislocation motion (e.g., in solid $^4$He \cite{iwasa,PBD}) can
affect the finite frequency, mechanical response of crystals
\cite{prb,joshy,he3prl}. For solid $^4$He the primary pinning
mechanism is the binding energy (between 300 mK and 1 K) between
dislocations and $^3$He impurities. In the zero temperature limit
all available binding sites around the dislocation network are
occupied by $^3$He. The density of isotopic impurities in the
vicinity of dislocations decreases at finite {\it T} due to the
thermally activated evaporation of $^3$He. Once freed of $^3$He, the
dislocation lines suffer from very few scattering processes and can
therefore vibrate freely and very effectively soften the crystal
\cite{joshy,day}. Such a phenomenon in H$_2$ crystals, which have
the same hexagonal-close-packed (hcp) structure and similar
dislocation densities ($\sim$10$^8$ cm$^{-2}$ \cite{h2dis}) as
$^4$He, could result in a temperature dependent resonant period well
below {\it T$_m$}. The interaction strength between dislocations and
HD impurities is $\sim$5 K and is much larger than for $^3$He in
solid $^4$He. However, for both singles and pairs of o-H$_2$
molecules the calculated binding energy is between 400 mK and 1 K
for different types of dislocations. One must therefore recognize
the potential importance of the concentration and mobility of
o-H$_2$ on $\tau$ even at milliKelvin temperatures.

Clearly, both singles and pairs can escape from potential wells
around dislocation lines within the sample if {\it T} is
sufficiently high (tens to hundreds of milliKelvin). However, two
other very important prerequisites are that these types of
impurities are actually present in the sample and that they are
mobile. For instance, in thermodynamic equilibrium the relative
population of singles is negligible below $\sim$300 mK
\cite{spt1,spt2}. At higher temperature their number increases, as
does their mobility. On the contrary, the population of pairs
increases only up to {\it T} $\sim$ 500 mK. Moreover, they are only
highly mobile below $\sim$100 mK. With this information in mind the
interpretation of the data in Fig.~\ref{fig:3} is as follows. Below
60 mK the o-H$_2$ configuration is quasi-static. Any small drifts in
$\tau$ are likely accounted for by the continual OP conversion
process. That is, the flipping of one spin can cause a triangle to
become a pair, which can diffuse. The o-H$_2$ impurities are
otherwise immobile. Each of the triangles and larger clusters serve
as pinning sites for dislocations. However, above 60 mK a finite
number of pairs is thermodynamically favorable (see
Fig.~\ref{fig:3}c), and thus the gradual breakup of clusters will
ensue. Since all of these pairs (both in-plane and out-of-plane
\cite{III}) are mobile for {\it T} $>$ 60 mK \cite{kran}, they need
not be bound to the dislocation network. This can lower the
effective elastic moduli. The sharp increase in $\Delta\tau$ at 60
mK is consistent with a growing number of propagating pairs. This
begins to level off above 100 mK as pair mobility drops, limiting
their escape rate from various binding sites. The $\Delta\tau$
observed in this temperature range will be smaller in samples with
higher ortho-concentrations since the crossover from network-pinning
to impurity-pinning \cite{iwasa,PBD,he3prl} occurs at higher {\it
T}. Above 400 mK the slope of $\Delta\tau$ in Fig.~\ref{fig:3}
increases again and is most likely due to the enhanced population of
isolated o-H$_2$ molecules in the sample. Since these singles become
more mobile between 300 mK and $>$ 1 K, they will continue to
successfully evaporate from dislocation lines. Thus, the crystal
should continue to soften with increasing {\it T} over this entire
temperature interval, as is observed.

There are several remarks to make concerning the expected magnitude
of $\Delta\tau$ stemming from changes to elastic moduli. First, one
clear conclusion based on FEM calculations is that only changes in
the shear (versus bulk) modulus could be reflected in resonant
period. This appears intuitive from the symmetry within the cell of
the open annulus TO, but is not obvious for the blocked annulus
cell. The nearly identical behavior expected from shear modulus
changes for both the open and blocked annulus configurations (see
Table~\ref{tab:1}) was already noted in \cite{prb} as strong
evidence for a supersolid $^4$He phase. In the present work the same
test not only serves as very solid evidence {\it against} a
supersolid H$_2$ phase, but demonstrates the qualitative agreement
between experiments and FEM calculations.

Second, the influence of {\it x} might be expected to oppose that
displayed in Fig.~\ref{fig:3}. That is, it is appropriate to ask
whether or not greater amounts of impurities produce larger period
shifts. In the simplest picture presented here $\Delta\tau$, mainly
depends on the difference in {\it G} for the two cases of ``all
dislocations pinned'' ({\it T} = 0 K) and ``all dislocations freely
vibrating'' (limited only by intersecting dislocation lines). This
quantity is independent of {\it x}, but depends on sample quality
(dislocation density). We actually observe a larger effect at lower
concentrations that, as mentioned above, probably reflects the lower
impurity-pinning crossover temperature in samples with lower
ortho-concentrations \cite{he3prl}, but also will be influenced by
the variations the dislocation network from sample to sample.

Third, although the model shown in Fig.~\ref{fig:6}a is the most
similar in design to the experimental TOs, the absolute accuracy of
$\Delta\tau_{FEM}$ values obtained from it is uncertain. For
instance, the difference between using it and using the simpler
versions in Figs.~\ref{fig:6}b and \ref{fig:6}c is a factor of
almost five in magnitude. This appears somewhat contrary to the fact
that small changes to the internal geometry (tested with the
simplified TOs) do not significantly alter $\Delta\tau_{FEM}$.
Taking the most dramatic result from Table~\ref{tab:1}
($\Delta\tau_{FEM}$ = 0.733 ns per 10\% decrease in {\it G}), we
find that between 0 and 200 mK (1 K) the shear modulus must drop by
20\% (68\%) to explain the 1.5 ns (5 ns) difference between the
H$_2$ and HD curves plotted in Fig.~\ref{fig:3}b. Some of the
discrepancy between experiment and FEM analysis could be accounted
for by viscoelasticity \cite{dorsey} and/or glassy behavior
\cite{balatsky}, which can be extracted from features in the
dissipation. Further, it may be that the additional components (Mg,
porous gold, etc.) within the cell are less coupled to the TO when
H$_2$ softens. This glue-like effect could enhance $\Delta\tau$. We
note that although we are unable to quantitatively simulate the
effect, it would be straightforward to repeat the experiment with a
very simple TO design.

The same physical picture can be applied to Fig.~\ref{fig:4}.
Relaxation times are extremely long below 60 mK since there are very
few mobile impurities. Above this temperature the slow dissociation
of clusters and the diffusion of pairs into some quasi-equilibrium
configuration will result in long relaxation times. The decrease in
{\it t$_O$} as {\it T} increases toward 200 mK could reflect the
increasing dissociation rate of clusters. As {\it T} is raised
further the motion of pairs will slow down and the relaxation
process will become dominated by the breakup into singles, their
liberation from dislocations, and their faster diffusion above
several hundred milliKelvin.

In TO experiments on $^4$He the dissipation peak (unlike the period
shift) can be entirely accounted for by glassy
\cite{balatsky,seamus} or viscoelastic \cite{dorsey} effects. Such
behavior is most likely linked to the influence of
$^3$He-dislocation interactions \cite{iwasa,PBD,he3prl} on the shear
modulus \cite{prb,day}. In H$_2$ there are at least two processes
that could cause dissipation for {\it T} $<$ 1 K. First, there is
the possible unbinding of o-H$_2$ molecules from dislocations, for
which we expect $\Delta${\it Q}$^{-1}$ to be centered around the
associated $\Delta\tau$. Second, there is the dissociation of
clusters, which should also affect the mechanical response of the
solid. By comparing Fig.~\ref{fig:3}b with Fig.~\ref{fig:5}, we find
that none of the ``peaks'' in dissipation are centered around large
$\Delta\tau$. For example, the peak at 200 mK occurs between the
period shifts at $\sim$80 and $>$ 300 mK. We speculate that the
dissociation of o-H$_2$ clusters, marked by large relative
populations of pairs and singles (see Fig.~\ref{fig:3}c), is
primarily responsible for the regions of enhanced dissipation.

%

\section{Conclusions}
\label{sec:6}

The mechanical response of solid hydrogen samples to extremely low
stresses has been measured using a torsional oscillator. A careful
analysis of the data, complemented by FEM calculations, confirm the
conclusion of \cite{h2prl} that the change in the TO resonant period
is not related to the onset of superfluidity in solid H$_2$. Rather,
$\Delta\tau$ is found to reflect changes in the shear modulus of the
sample. The most prominent shift occurs between 60 and 180 mK and
most likely corresponds to the breakup of large o-H$_2$ clusters
into triangles and pairs of molecules. Further softening of the
solid occurs at higher temperature as clusters gradually separate
into isolated molecules. The temperature intervals over which the
unclustering takes place is complemented by enhanced internal
dissipation.

\begin{acknowledgements}
Thanks to P. J. Polissar and C. Turich for assistance with mass
spectrometry. We are particularly grateful to H. Meyer for his
advice, numerous communications on solid hydrogen and warm
encouragements.
\end{acknowledgements}



\end{document}